\documentclass[conference]{IEEEtran}
\IEEEoverridecommandlockouts
\usepackage{cite}
\usepackage{amsmath,amssymb,amsfonts}
\usepackage{algorithmic}
\usepackage{graphicx}
\usepackage{textcomp}
\usepackage{xcolor}

\usepackage{booktabs} % For formal tables

\def\BibTeX{{\rm B\kern-.05em{\sc i\kern-.025em b}\kern-.08em
    T\kern-.1667em\lower.7ex\hbox{E}\kern-.125emX}}

\newcommand{\bi}{\begin{itemize}}
\newcommand{\ei}{\end{itemize}}
\newcommand{\be}{\begin{enumerate}}
\newcommand{\ee}{\end{enumerate}}
\newcommand{\bv}{\begin{verbatim}}
\newcommand{\bd}{\begin{description}}
\newcommand{\ed}{\end{description}}

\begin{document}

\title{Detecting the Insider's Threat with Long Short Term Memory (LSTM) Neural Networks\\
{\footnotesize 
%\textsuperscript{*}Note: Sub-titles are not captured in Xplore and should not be used
}
\thanks{Identify applicable funding agency here. If none, delete this.}
}

\author{
\IEEEauthorblockN{ Eduardo Lopez}
\IEEEauthorblockA{\textit{DeGroote School of Business} \\
\textit{McMaster University}\\
Hamilton, ON, Canada \\
lopeze1@mcmaster.ca}
\and

\IEEEauthorblockN{ Kamran Sartipi}
\IEEEauthorblockA{\textit{Department of Computer Science} \\
\textit{East Carolina University}\\
Greenville, NC, USA\\
sartipik16@ecu.edu}
}

\maketitle

%------------------------------------------ BEGIN SECTIONS ------------------------------------------

%%%             Important Tips in Writing a Paper
%%%    Write "CONSICE" and "SPECIFIC" and "CLEAR" and "COHERENT".
%%%
%%%%%%%%%%%%%%%%%%%%%%  ABSTRACT %%%%%%%%%%%%%%%%%%%%%%%%%%%%%%%%%
% GUIDES:
% - Provide a summary of the paper in general terms and with a motivating theme.
% - Start with one or two sentences to pose an existing problem and briefly explains how
%   the proposed technique in the paper will solve it.
%%%%%%%%%%%%%%%%%%%%%%%%%%%%%%%%%%%%%%%%%%%%%%%%%%%%%%%%%%%%%%%%%%
\begin{abstract}
%\vspace{-3mm}
%
Information systems enable many organizational processes in every industry. The efficiencies and effectiveness in the use of information technologies create an unintended byproduct: misuse by existing users or somebody impersonating them - an insider's threat. Detecting the insider's threat may be possible if thorough analysis of electronic logs -- capturing user behaviors -- takes place. However, logs are usually very large and unstructured, posing significant challenges for organizations. In this study, we use deep learning, and most specifically Long Short Term Memory (LSTM) recurrent networks for enabling the detection. We demonstrate through a very large, anonymized dataset how LSTM uses the sequenced nature of the data for reducing the search space and making the work of a security analyst more effective.

%\textcolor{red}{NOTE: \\ At the top of each provided latex file there is a guideline for the corresponding section. * Comment this *. \\
%The IEEE Sample Latex  instructions are in file "8.IEEEconference\_Smple.tex".   It consists of examples of  Latex Instructions to prepare a paper.  To activate it you should un-comment  the "8.IEEEconference\_Smple.tex" tex file.
%}

\end{abstract}


%%%%%%%%%%%%%%%%%%%%%%%%%%%%%%%%% INTRODUCTION %%%%%%%%%%%%%%%%%%%%%%%%%%%%%%%%%
% GUIDES:
% - The major goal is to motivate the reader to continue reading the paper. 
% - Introduce the problem to be addressed by the paper and the related scientific field(s) 
% - Describe briefly the proposed approach in the paper. 
% - Provide a list of contributions in the paper. 
% - Introduce the titles of the paper's sections (paper's structure)
%%%%%%%%%%%%%%%%%%%%%%%%%%%%%%%%%%%%%%%%%%%%%%%%%%%%%%%%%%%%%%%%%%%%%%%%%%%%%%%%
\section{Introduction}
%\vspace{-3mm}
Most organizations use Information Technology (IT) for the enabling of their processes. The higher productivity and process efficiencies gained with the use of information systems come with a byproduct: the risk of individuals misusing technology. We define the insider's threat as the misuse or non-authorized use of information systems by users (or somebody impersonating them). Detecting the insider's threat is a difficult task, with the ever-increasing number, complexity and inter-connectivity of information systems enabling many activities in an organization. Although the user actions are usually stored in electronic files or logs, it is difficult to detect malicious activity just by reading them. Logs are usually very large files, and contain myriad data points that are either irrelevant to the analysis or normal operation of the system.

The experiments described in this study aim to demonstrate how the use of deep learning, and more specifically Long Short Term Memory (LSTM) neural networks, can be used for an effective detection of the insider's threat. The proposed approach is aligned with the conditions found in real world settings, where very large datasets and finite resources are prevalent - with a focus on short computational time and probabilistic output. This study contributes to extant research by articulating a practical approach that enables a structured detection within constrained of time and resources. 

This paper is organized as follows: in Section \ref{framework} we depart from neural networks foundations and move towards more advanced concepts such as deep learning and recurrent neural networks.  Section \ref{approach} describes how we architect detection through the selected hardware and software, followed by the experiments in Section \ref{Experimentation}. We close the document with our conclusions and suggested research streams in Section \ref{conclusion}.  

%\input{2.RelatedWork}
%%%%%%%%%%%%%%%%%%%%%%%%%%%%%%%%% Background %%%%%%%%%%%%%%%%%%%%%%%%%%%%%%%%%
% GUIDES:
% Describe the required background about methodology, technology, or tools
% that are needed to follow and understand the paper. 
%%%%%%%%%%%%%%%%%%%%%%%%%%%%%%%%%%%%%%%%%%%%%%%%%%%%%%%%%%%%%%%%%%%%%%%%%%%%%%

\section{Background and Related Work}  \label{framework}
%\vspace{-3mm}
% 
The experiments described in this study are anchored in computational and mathematical models. In this section we explain the theoretical framework used and how it is applied towards the detection of an insider's threat.
\color{black}
\subsection{Neural networks foundations}

Neural networks are computational models used for classification and prediction.  On the most basic form, a neural network takes a set of inputs, or predictors, performs some calculations on it and provides an output or outputs as the result of the process.  A simple neural network with one node takes the inputs and generates the output based on weights that it assigns to each of the inputs.

\begin{figure}[h]
	\centering
	\includegraphics[width=0.7\linewidth]{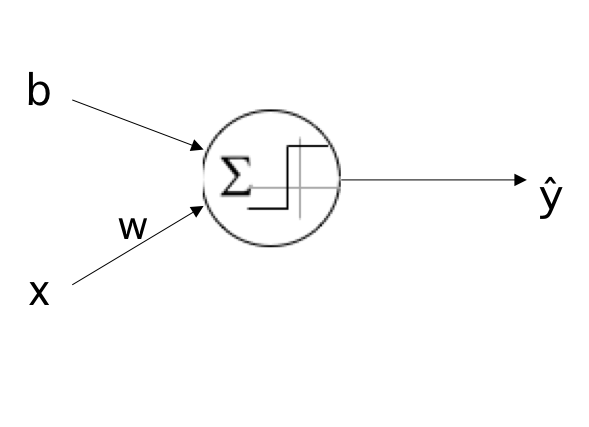}
	\caption{A single neuron processing model.}
	\label{fig:SingleNeuron}
\end{figure}

In Figure \ref{fig:SingleNeuron}, there are two inputs: x to denote a vector of observations and b as the bias. The neural network assigns the weights w and sums the resulting numbers. The results of this operation are then entered into an activation function, which is used to introduce non-linearity into the output. There are three typical activation functions found in the literature. They are depicted in Figure \ref{fig:nnactivation}. 

\begin{figure}[h]
	\centering
	\includegraphics[width=1\linewidth]{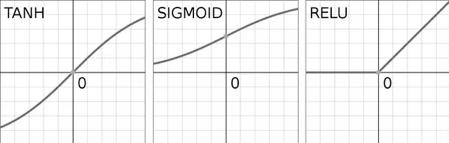}
	\caption{Activation Functions used in Neural Networks}
	\label{fig:nnactivation}
\end{figure}

If the activation function we use is the sigmoid (see Figure \ref{fig:nnactivation}), the function calculation would be as follows:

\begin{equation}
\hat{y}= \sigma (w*x+b)
\end{equation}

where $\hat{y}$ is the predicted value. The $\sigma$ function is defined as:
\begin{equation}
\sigma(z)=  \frac{1}{1+e^{-z}} 
\end{equation}

Interconnecting multiple neurons and arranging them in layers is illustrated in Figure \ref{fig:nn3neuron}. 

\begin{figure}[h]
	\centering
	\includegraphics[width=1\linewidth]{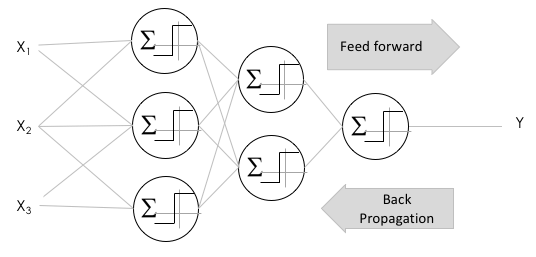}
	\caption{Neural Network with three layers.}
	\label{fig:nn3neuron}
\end{figure}

For the neural network we define a loss function for \textit{one} observation to measure how close the prediction $\hat{y}$ is to the actual value y. 

\begin{equation}
L(\hat{y},y)= -(y*log(\hat{y})+(1-y)*log(1-\hat{y}))
\label{lossFunction}
\end{equation}

Equation \ref{lossFunction} is a quantified metric on how close the prediction is to the actual value observed by the neural network.  
For all observations (i.e. training data with \textit{m} records) we define the cost function \textit{J} as follows:

\begin{equation}
J(w,b)=  \frac{1}{m}*\sum_{i=1}^{m}L(\hat{y}_i, y_i)
\label{lossFunction}
\end{equation}

A three-dimensional illustration of Equation \ref{lossFunction} is depicted in Figure \ref{fig:lossfunctionnn}.

\begin{figure}[h]
	\centering
	\includegraphics[width=0.7\linewidth]{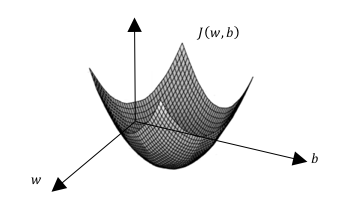}
	\caption{Loss function J mapped against w and b.}
	\label{fig:lossfunctionnn}
\end{figure}

The closest this number is to zero, the more accurate the model is at predicting results with the given training dataset. Thus, the minimization of Equation \ref{lossFunction} is a mathematical objective which is performed through an optimization algorithm called \textit{gradient descent}.  Graphically this can be seen as traversing the function in Figure \ref{fig:lossfunctionnn} towards its global minimum.

The slope of Equation \ref{lossFunction} is calculated as 
\begin{equation}
\frac{\partial J(w,b)}{\partial w}
\end{equation}

The next values for w and b are calculated on each epoch (forward and backward propagation) as follows:
\begin{equation}
w:= w- \alpha*\frac{\partial J(w,b)}{\partial w}
\end{equation}
\begin{equation}
b:= b - \alpha* \frac{\partial J(w,b)}{\partial b}  
\end{equation}
 
Where $\alpha$ is known as the learning rate. The process by which the gradient descent takes place happens through a feed forward of data, with back propagation recalculating parameters. 
The features are organized in arrays that the neural network can understand (i.e.,  \textit{tensors}). A tensor can be defined as a container of data which can be considered as a "bucket of numbers" \cite{daniel_jeffries_learning_2019}. Tensors are the fundamental data structures in deep learning models \cite{noauthor_tensors_2018}. An essential tensor attribute is its shape which conveys how the numbers are stored. A 0-dimensional (i.e., 0D) tensor is a single number, a scalar. A 1D tensor is a vector of numbers. 2D tensor is a matrix, with a 3D tensor being a cube.  Tensors traverse the multiple layers forward and backward as the estimation of the parameters take place.

\subsection{Deep learning and gated neural networks}
\label{machine learning}
A deep network is defined as a neural network in which multiple layers exist. As more layers are added, the more difficult it becomes to calculate the right weights; the signal becomes 'weaker'.  This is sometimes referred to as the vanishing gradient \cite{yoshua_bengio_patrice_simard_paolo_frasconi_learning_1994}. This is an important challenge given that the weights are initially randomly assigned. It used to be impossible to parametrize a deep learning network with back propagation.  The tools and techniques used in addressing this challenge are labelled as deep learning.

Deep learning can be described as algorithms for feature detection.  This was a breakthrough made by some researchers in the field - more notably Hinton from the University of Toronto - in 2006.  The basic concept is that each layer in the neural network learns features.  More specifically, the first layer learns primitive features like the edge of an image.  Once it has learned those features, they are fed to the next layer where more sophisticated features are recognized.  The process allows for the deep network to recognize increasingly complex features in the data \cite{Lecun2015d}.

A milestone was achieved in October 2012, when Hinton announced that two of his students had invented software that recognized objects with double the accuracy of their next competitor in the Artificial Intelligence challenge called ImagiNet.  From that point on, significant progress has been made in areas as varied as gaming, Natural Language Processing and image labeling.

The neural networks described previously are typically used with no conception of time or sequence dependency. However, it is quite common that the values depend on preceding values. A good illustration of this is Natural Language Processing. Predicting the last word in the phrase "I live in France, I speak ..." may be considered a classical case where the data in the sequence is essential to the prediction. This kind of problems can be approached by using neural networks that consider the preceding states, i.e., Recurrent Neural Networks or RNN. Conceptually, an RNN looks at preceding states for the output of the current observation. This is illustrated in Figure \ref{fig:rnn}, with the dashed arrows on the unrolled RNN representing the preceding states flow.

% TODO: \usepackage{graphicx} required
\begin{figure}[h]
	\centering
	\includegraphics[width=1\linewidth]{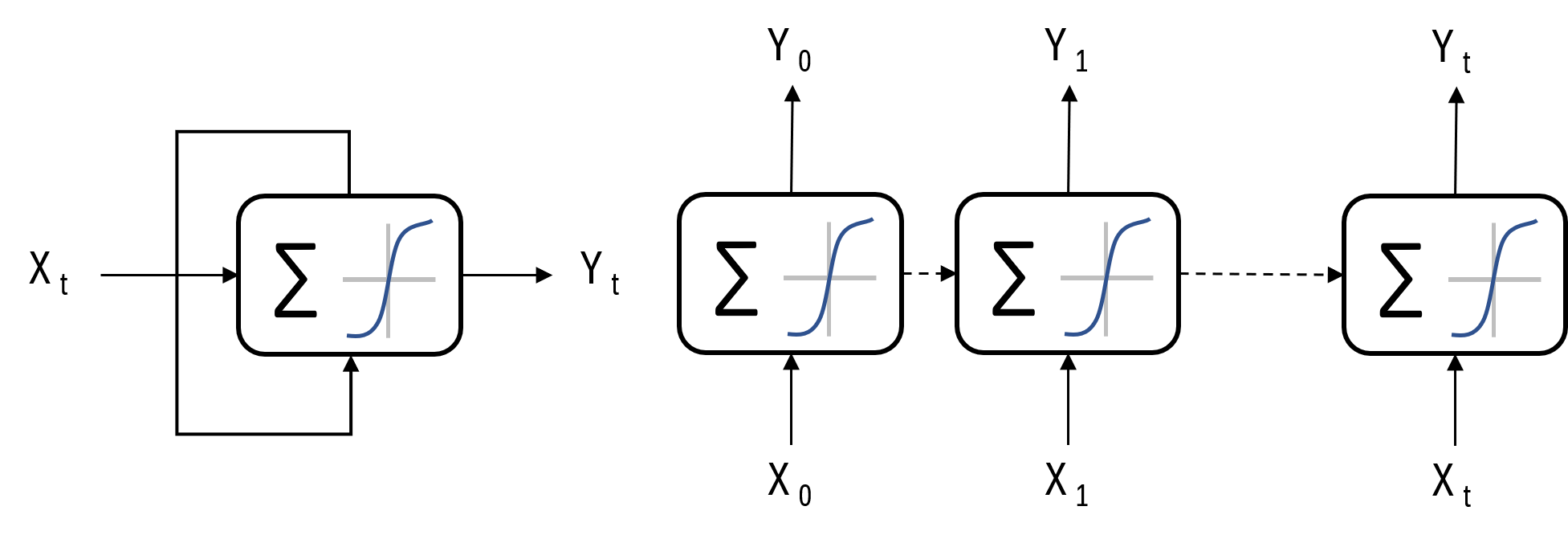}
	\caption{Recurrent Neural Networks}
	\label{fig:rnn}
\end{figure}

A fundamental principle is raised with RNN: how far in the past should the network reach when assessing an observation? The architecture depicted in Figure \ref{fig:rnn} will suffer from short-term memory. As information gets passed to the next layer, early information may be reduced or vanish entirely. To address this issue, new designs were implemented, such as Long Short Term Memory (LSTM) \cite{sepp_hochreiter_long_1997} and Gated Recurring Units (GRU) \cite{chung_empirical_2014}. 

LSTM neurons use a mechanism for regulating the flow of information, namely gates \cite{nguyenIllustratedGuideLSTM2019,UnderstandingLSTMNetworks}.The gates enable LSTM to learn what data in the sequence is important, and what can be discarded. An illustration of LSTM is depicted in Figure \ref{fig:lstm}. The LSTM design uses two distinct activations: tanh and sigmoid, as depicted previously in Figure \ref{fig:nnactivation}. 

% TODO: \usepackage{graphicx} required
\begin{figure}[h]
	\centering
	\includegraphics[width=1\linewidth]{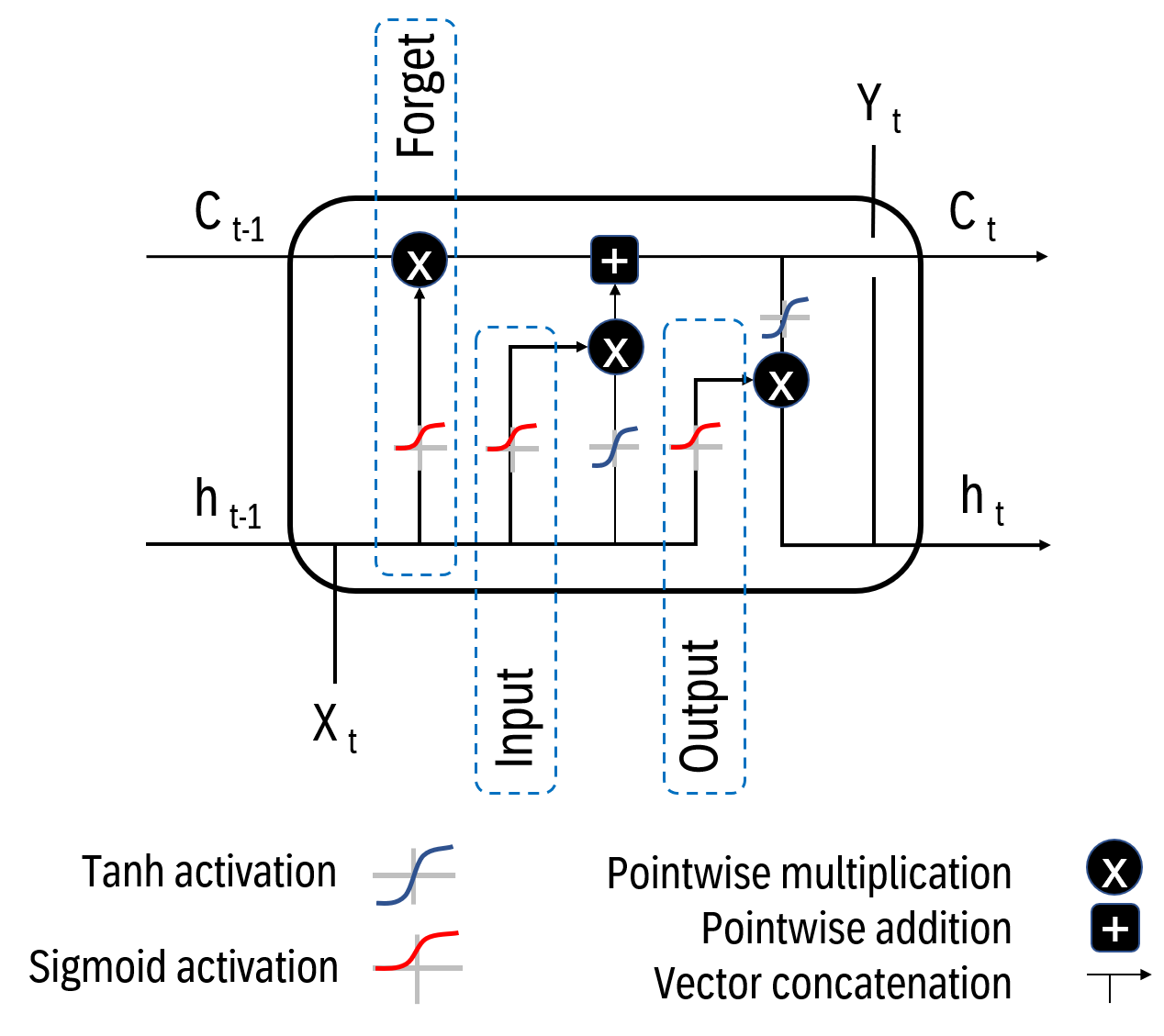}
	\caption{Long Short Term Memory - gates and operations}
	\label{fig:lstm}
\end{figure}

The first step in the processing is the vector concatenation of the preceding hidden state ($h_{t-1}$) and the input to the LSTM unit ($x_{t}$). This tensor is then multiplied by the weights matrix $W_{f}$, added the bias vector $b_{f}$, and  passed through a Sigmoid activation function $\sigma$.  The resulting vector is then elementwise multiplied by the preceding cell state $C_{t-1}$. The set of operations is referenced as the \textit{forget} gate, with the equation stated below.

\begin{equation}
f_{t}=(\sigma(W_{f}.[h_{t-1},x_{t}] + b_{f}))*C_{t-1} 
\end{equation}

%Where $W_{f}$ represents the weights assigned and $b_{f}$ the bias.
In the next gate, we depart from the same input $h_{t-1},x_{t}$ and move them through a Sigmoid activation:

\begin{equation}
i_{t}=\sigma(W_{i}.[h_{t-1},x_{t}] + b_{i}) 
\end{equation}

And a Tanh activation:
\begin{equation}
\overline{C}_{t}=tanh(W_{c}.[h_{t-1},x_{t}] + b_{c}) 
\end{equation}

$\overline{C}_{t}$ is multiplied with $i_{t}$ and added to the cell state coming from the forget gate. This value is the new cell state and is calculated with:
\begin{equation}
C_{t}=f_{t}*C_{t-1}+i_{t}*\overline{C}_{t} 
\end{equation}

The new cell state includes what information we want to forget, as well as the information that is deemed important.

The final stage is the output gate. The first value calculated uses a Sigmoid activation:
\begin{equation}
o_{t}=\sigma(W_{o}.[h_{t-1},x_{t}] + b_{o}) 
\end{equation}

Which is then multiplied by the cell state put through a Tanh activation:
\begin{equation}
h_{t}=o_{t}*tanh(C_{t})
\end{equation}

Many different architectures have evolved from this LSTM concept. Multiple LSTM stacked, or using bi-directional LSTM have produced better results that single LSTM \cite{gravesHybridSpeechRecognition2013}.

\subsection{Keras}
There are multiple deep learning frameworks that implement the concepts discussed above. TensorFlow \cite{noauthor_tensorflow_2020} is arguably one of the best known framework, originally created by Google. It works with a static computation graph, where re-training is required when making changes to the architecture. A higher-level API can be used on top of TensorFlow, making deep learning development and prototyping more efficient. The best known is Keras, which we select for this study.

A Keras model is composed of multiple layers connected in sequence that progressively process tensors \cite{noauthor_home_nodate}. The followings are some of the layers available.
 
\paragraph{Dense layer} These are the most common deep learning layers. In a typical Keras model, the dense layers uses two attributes: the {\it input shape} and the {\it number of units}. In its simplest form, the input shape is the dimension of the feature vector. In Figure \ref{fig:nn3neuron}, the input shape is 3 and the number of units of the first dense layer is 3 (since there are three neurons). As more dense layers are added to a Keras model, the shape of the tensor is not required anymore since it was defined in the first layer. In Figure \ref{fig:nn3neuron} the second dense layer only has, as arguments, the number of units or neurons (i.e., 2). 

 \paragraph{Dropout layer} A common issue with deep neural networks is overfitting. Using dropout is a technique that can be used to avoid this situation. It consists of randomly dropping units in the neural network during training \cite{srivastava_dropout_2014}. To add a dropout layer in Keras, a rate is used as parameter, and that pertains to the fraction of units that are randomly set to 0.
 
 \paragraph{Embedding layer} Frequently used in Natural Language Processing (NLP), an embedding layer converts positive integers into dense vectors. This is suitable to situations in which the vocabulary set is very large, and a lower-dimensional vector is more efficient for the processing of the dataset.
 
 \paragraph{Recurrent layer} There are multiple implementation of recurrent networks in Keras, from the simple RNN to gated units such as Gated Recurrent Units (GRU) and Long Short Term Memory (LSTM) networks. LSTM is implemented in alignment with the work by Hochreiter et al. in 1997 \cite{hochreiter97:LSTM}. 
 Many other Keras layers exist, including pooling and convolutional layers, which can be used for specific applications in text and image processing.

\subsection{Measuring success}
A foundational element in the use of any prediction or classification problem pertains to how it is compared against other models or a given baseline. The approach this study makes to the research challenge -- identifying the insider's threat through anomaly detection -- is ultimately a classification exercise: given historical sequenced data, is the observation normal or abnormal (two-class). A classical way of evaluating performance of a classifier is through a confusion matrix \cite{shmueli2017data}, depicted in Figure \ref{fig:confusionmatrix}.

\begin{figure}[h]
	\centering
	\includegraphics[width=0.6\linewidth]{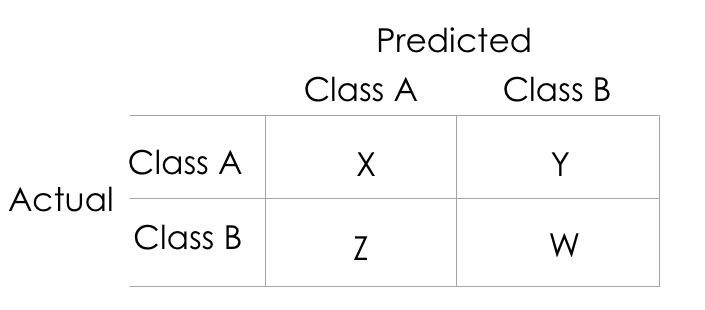}
	\caption{Confusion matrix for a binary classifier.}
	\label{fig:confusionmatrix}
\end{figure}

$X+Z$ is the number of observations that the classifier predicts as class A (i.e. probabilities higher than the threshold of 0.5). \textit{X} are the correctly predicted, and \textit{Z} were incorrect.  \textit{sensitivity} is the ability of the model for correctly predicting a class (true positives), and {\it specificity} of the classifier is its ability to correctly predict the other class, i.e., class B (true negatives). 
\begin{equation}
	sensitivity = true\_positive\_rate = \frac{X}{X + Y}
\end{equation}

\begin{equation}
specificity = true\_negative\_rate = \frac{W}{Z + W}
\end{equation}

Sensitivity and specificity of a classifier change as different probability thresholds are used. A perfect classifier has a sensitivity of 100\% and a specificity of 100\%. Plotting the performance of a binary classifier is possible through a ROC (Receiver operating characteristic) curve \cite{shmueli2017data}, as illustrated in Figure \ref{fig:exampleroc}.

\begin{figure}[h]
	\centering
	\includegraphics[width=1\linewidth]{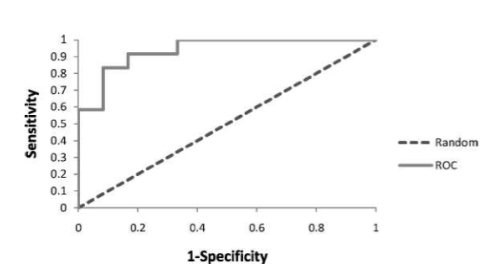}
	\caption{Example ROC curve for a binary classifier.}
	\label{fig:exampleroc}
\end{figure}

A very good classifier that separates correctly the two classes would appear on the upper left point [0,1]. A very bad classifier would be equivalent to a random draw, with a curve close to the diagonal. The Area Under the Curve (AUC) is usually used to represent the quality of a given classifier as captured in a ROC curve \cite{shmueli2017data}.

%%%%%%%%%%%%%%%%%%%%%%%%%%%%%%%%% APPROACH %%%%%%%%%%%%%%%%%%%%%%%%%%%%%%%%%
% GUIDES:
% - Describe the main technical aspect and contributions of the paper.
% - Be as specific and clear as possible.
% - Producing an overview of the approach that summarizes all the steps in one place
%   would help the reader to understand and follow the rest of the paper.
% - Organizing the description of the approach into steps that are related to the overall view
%   of the approach will ease your task and assist the reader to follow the subjects.
%%%%%%%%%%%%%%%%%%%%%%%%%%%%%%%%%%%%%%%%%%%%%%%%%%%%%%%%%%%%%%%%%%%%%%%%%%%%%

\section{Approach} \label{approach}
%\vspace{-3mm}
%
This section articulates the context and decisions influencing the design of the experiments.
\color{black}
\subsection{Principles and constraints}
Detecting the misuse of information systems by mining the data contained in electronic logs is a challenging task. It is specially difficult if there is no "signature" or baseline of what is considered a normal behavior. Experiments such as the ones described in this work must follow as closely as possible what exists in reality. Some of the constraints worth mentioning are:
\subsubsection{No normality baseline} Data available in logs do not have labels. The records just describe what is happening in information systems, but is not classified as normal or abnormal. This is a critical constraint we use as it is in alignment with reality in which data is not pre-labeled.
\subsubsection{Finite training and detection time} The identification of an insider's threat taking place is necessarily a time-bound activity. Detecting the misuse of an information system may be impractical if it needs too long to take place.
\subsubsection{Very large datasets} Electronic logs are usually very large datasets. The volume of information -- irrespective to the relevancy of the analysis -- demands tools (hardware and software) that is specialized and powerful.
\subsubsection{Probability-based} Detecting an insider's threat is not an exact process. The objective is to improve the probability of a human resource -- e.g. a security analyst -- effectively and efficiently detecting the threat. A corollary to this is that false positives or true negatives are acceptable as long as the detection process is enhanced.
\subsubsection{Inherent characteristics in the data}
The algorithms and computational statistics to be used must take advantage of the nature of the data: whether is/is not parametrized, spatial or sequential/temporal, categorical or numerical, etc.
\subsection{Dataset}
The paucity of datasets in the are of cybersecurity has been a known issue for researchers. However, with the explosion of digital data availability and remarkable interest in machine learning, this is no longer the case.  The Los Alamos national laboratory in the United States has published a remarkably rich dataset that can enable myriad research projects \cite{akent-2015-enterprise-data}. The dataset is an anonymized, multi-device log captured through the monitoring of servers, workstations and network devices during 58 days of uninterrupted activity. The data includes more than 12,000 users, using more than 17,000 computing devices. Furthermore, the data includes actions taken by a 'red team', i.e., insider threat attacks in the network.

There are multiple log files that can be analyzed. Figure \ref{fig:losalamos} depicts the relative sizes. As can be observed, the \textit{authentications} log contains more than 1 billion records. It captures the timestamp of the event, as well as the user and computers involved. 

% TODO: \usepackage{graphicx} required
\begin{figure}[h]
	\centering
	\includegraphics[width=1\linewidth]{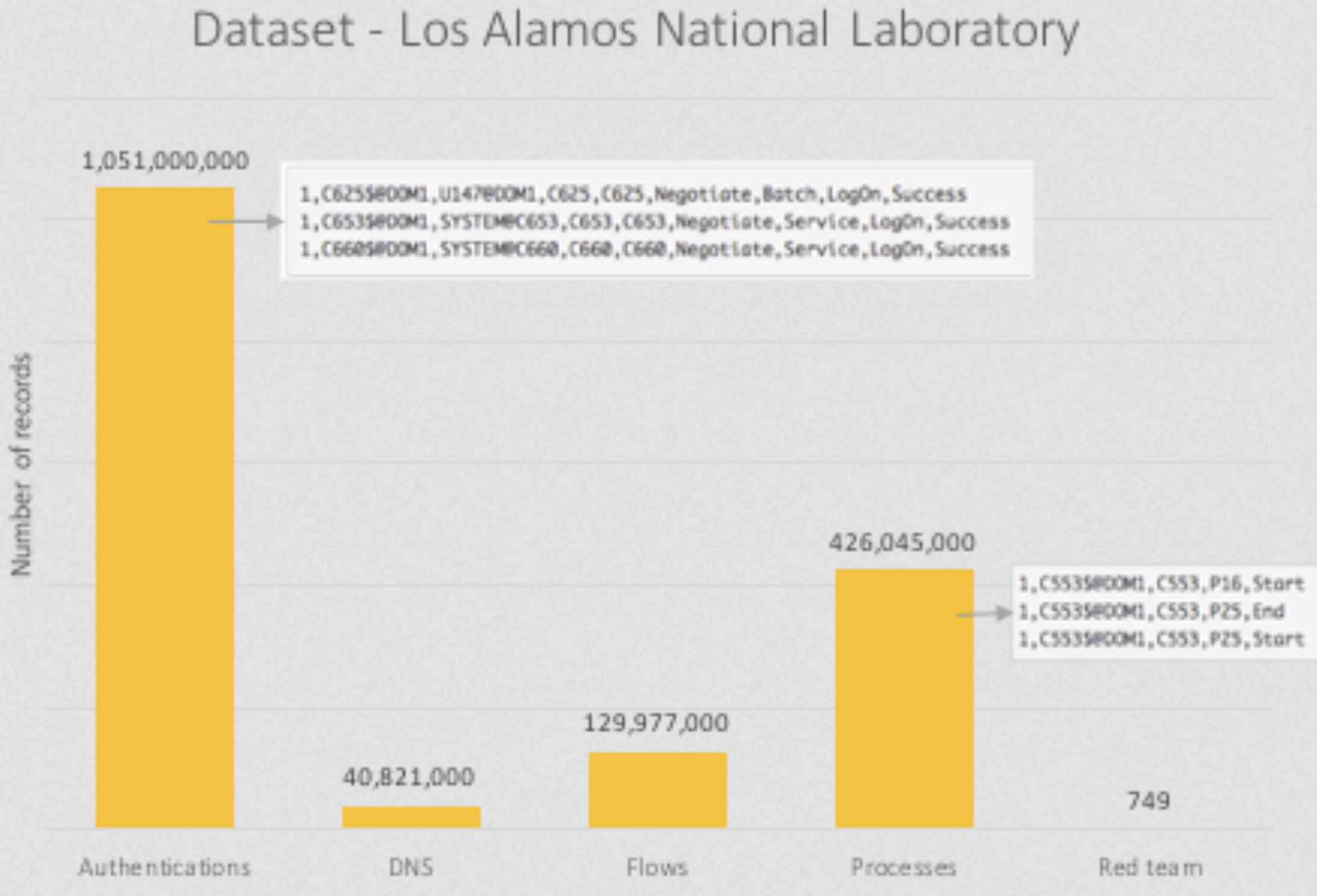}
	\caption{Los Alamos National Laboratory - log files created over 58 days of monitoring.}
	\label{fig:losalamos}
\end{figure}

For the purposes of this study, we select the authentications file as our research focus. 

\subsection{Technology Stack}

The processing of a very large dataset requires specialized technology tools. The dataset contains approximately 1.6 billion records stored in flat files. We select Apache Spark as the data analytics engine for performing the pre-processing tasks: ensuring the data is suitable for consumption by machine learning models. 

Apache Spark \cite{ApacheSp6:online} enables data analysis through multiple modules such as Spark SQL and MLlib. As of the time of this study, deep learning can be implemented in Apache Spark with the use of a dedicated package \cite{noauthor_databricks/spark-deep-learning_2020} or through other methods that make use of TensorFlow and Keras \cite{vazquez_deep_2018}. However, given our intent on fast prototyping and experimentation, we select Knime as the front-end tool utilized in this study. Knime \cite{KNIMEOpe0:online} is a modular data pipelining tool that enables implementation of machine learning models with minimal coding. Both Knime and Apache Spark (in its commercial form: Databricks) are highly rated by industry, as captured in the Gartner magic quadrant for data science and machine learning platforms \cite{noauthor_gartner_nodate}.

In terms of the hardware infrastructure, a purpose-built DGX-1 machine by NVIDIA is used. The DGX runs the software described with Ubuntu server as the operating system.

Figure \ref{fig:architecture} depicts the overall approach, with the hardware and software described. Please note the different processes to be performed, and the hardware and software artifacts that enable them.

% TODO: \usepackage{graphicx} required
\begin{figure}[h]
	\centering
	\includegraphics[width=1\linewidth]{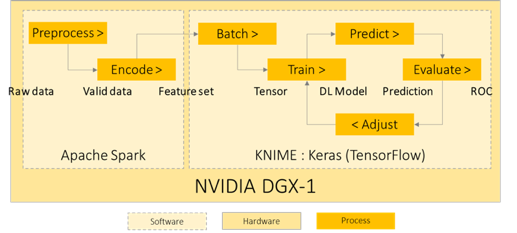}
	\caption{Architecture for detection of the insiders' threat - hardware and software}
	\label{fig:architecture}
\end{figure}

%%%%%%%%%%%%%%%%%%%%%%%%%%%%%%%%% CASE STUDY %%%%%%%%%%%%%%%%%%%%%%%%%%%%%%%%%
% GUIDES:
% - Describe one or more case sudies as proof of concept to support the proposed
%   approach and methodology described in the paper. 
% - The case studies can be either simulations or experimentations with a prototype tool. 
% - The results should be demonstrated using figures, tables or graphs. 
%%%%%%%%%%%%%%%%%%%%%%%%%%%%%%%%%%%%%%%%%%%%%%%%%%%%%%%%%%%%%%%%%%%%%%%%%%%%%%%
\section{Experimentation} \label{Experimentation}
This section will be divided in segments in alignment with the processes depicted in Figure \ref{fig:architecture}, including: pre-processing, encoding, batching, training, prediction and evaluation.

\subsection{Pre-processing} The raw data is produced from the multiple devices that are connected to the network. The data is anonymized and collected in a CSV (Comma-Separated Values) format. This first step is depicted in Figure \ref{fig:rawdatacreation}.

% TODO: \usepackage{graphicx} required
\begin{figure}[h]
	\centering
	\includegraphics[width=0.7
	\linewidth]{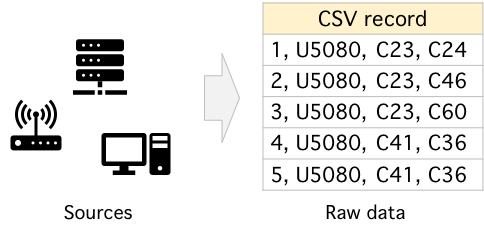}
	\caption{Creation of raw data from devices, anonymized into a CSV file}
	\label{fig:rawdatacreation}
\end{figure}

The raw data needs to be cleansed and invalid records need to be either inputted or deleted (following list-wise deletion best practices). The data must be transformed from string into a suitable type that permits further manipulation. The information provided in an authentication record includes the user involved in the event prefixed with the letter "U". It also includes both the source computer and the destination computer, both prefixed with the letter "C". The numbers (1 to 5) in the left column of Figure \ref{fig:rawdatacreation}  indicate the seconds in which the events took place.

We proceed to inject into the dataset the ground truth, or known records capturing the interactions from the red team. This information is going to become essential when we measure the success in identifying the insiders' threat. 

It is important to note that this information is only used to assess the quality of the detection, but the action taken to identify the red team do not make use of this data, in alignment with conditions found in real-world applications where there is no signature pre-defined. In other words, the approach taken in this study is unsupervised learning -- the deep learning model architected does not receive which records are normal and which ones are produced by the red team.

A last, but important pre-process performed is the determination of the date baseline. The timestamp value in the raw data is a single scalar number capturing the number of seconds since the start of the monitoring. Analyzing the relative volume of data at different points in time, it is possible to identify the day/afternoon/evening patterns, as well as weekdays/weekends. We use this information to assign an absolute date timestamp to each record, with the very first one taking place the 1-Jan-2018.

\subsection{Encoding}
We now proceed to define an \textit{event}. For the purposes of this analysis an event is the representation of the unequivocal behavior that the user is performing at any given time. A record in the authentications log includes the event that takes place, and the timestamp associated with it.

As it was discussed in Figure \ref{fig:rawdatacreation}, the raw record comes with the timestamp, user, source computer and destination computer. A simple, yet complete event will, therefore, include these three elements as U5080, C23, C24, and it represents user U5080 authenticating from computer 23 to computer 24.

Once the events are generated, we proceed to encode them in a data type that is accepted by a computational model. In this study we refer to this task as \textit{dictionarizing}. Every unique event is assigned an integer number. Authentications that happen at different points in time may be for a common event -- a user may login in the morning, log off at noon, and then log back in for the afternoon. 

Figure \ref{fig:encoding} depicts the encoding that takes place. In the illustration there are 5 records, with a repeated event that takes place at 10am and 11am with the index number 4.

% TODO: \usepackage{graphicx} required
\begin{figure}[h]
	\centering
	\includegraphics[width=0.7
	\linewidth]{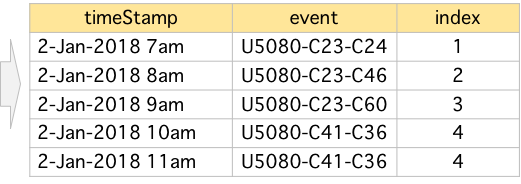}
	\caption{Encoding the pre-processed data.}
	\label{fig:encoding}
\end{figure}

It is important to note that there is a further step in the encoding that is not described in depth, but nonetheless takes place. The representation of each index uses one-hot encoding when inputted into the deep learning model. This process produces sparse vectors -- long arrays of zeroes (0) with a few ones (1). For simplicity and illustration, we do not include the one-hot encoding in the depictions for this work.

The encoding we have performed translates a categorical value (e.g. U5080-C23-C24) into a number that can be used by the mathematical model. The total number of unique events in the dataset is referred to as the vocabulary size. This encoding does not capture the relationships between the different events. The index chosen is arbitrary -- usually following some convention such as alphabetical order. It is possible to index the data with meaningful numbers -- for example by performing feature engineering to calculate word embeddings from the data. For simplicity purposes, we continue using simple indexing in these experiments. 

\begin{table}[h!]
	\centering
	\begin{tabular}{||c|c|c||} 
		\hline
		User & Vocabulary size & Highest event frequency \\ [0.5ex]
		\hline\hline
		U737 & 812 & 1,905 \\
		\hline
		U3005 & 182 & 3,102  \\
		\hline
		U213 & 100 & 2,875 \\
		\hline
		U1581 & 7 & 19 \\
		\hline
		
	\end{tabular}
	\vspace{3mm}
	\caption{Tensor characteristics of selected users}
	\label{tab:tensorUsers}
\end{table}

Table \ref{tab:tensorUsers} displays two key indicators for four users. Vocabulary size (i.e. how many different events exist) and the highest frequency for an event. As can be observed, user U737 has a larger quantity of unique events, whereas users U3005 and U213 have a significantly lower number of potential events that are more common (i.e. larger frequency). Further analysis of the behavior is possible, but is considered out of scope for the experiments in this study.

\subsection{Creation of batches}

The feature set produced in the encoding stage shall now be converted into a data structure that is typical of deep learning, i.e., tensors. A tensor is a multi-dimensional array used as the foundational data structure in deep learning \cite{noauthor_tensors_2018}. In a deep learning network, a tensor is moved from layer to layer as processing happens. The creation of batches produces the tensors that are fed into the model.

For the creation of the tensors, some design decisions are required. Arguably the most fundamental design decision is about the sequenced nature of the data. As can be observed in the dataset, the information follows a temporal order. The events defined in the encoding stage exist in a sequence that allows for the prediction of subsequent events. 

A second essential design decision is how far one needs to look into the sequence in order to make the predictions. In other words, does a sequence of \textit{x} events effectively enable prediction of \textit{y} events. To maintain a parsimonious model, we select \textit{y} to be 1. In practical terms this means that our model is a many-to-one, as it predicts the authentication that takes place based on the preceding \textit{y} events. 

The process is illustrated with user U5080. 
Figure \ref{fig:batcheddata} top shows 6 (un-encoded) authentication records, with 4 different indexed events (a vocabulary size of 4) and a window of 3 events in the past. 
Figure \ref{fig:batcheddata} bottom depicts the creation of the tensors with encoded events into numbers. 

% TODO: \usepackage{graphicx} required
\begin{figure}[h]
	\centering
	\includegraphics[width=1\linewidth]{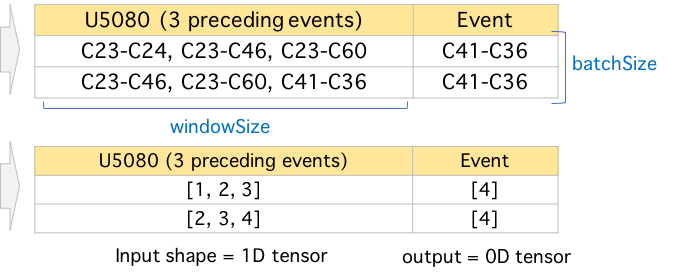}
	\caption{Tensors created based on window of 3 events.}
	\label{fig:batcheddata}
\end{figure}

As it can be observed, this is a many-to-one example, where the three precedent events are used to predict the fourth event. The vocabulary size means that the sparse vector representing event \textit{2} is a 1D-tensor with a 1 in the $2^{nd}$ position, and zeroes in all others. In addition to this, deep learning networks are trained with groups or batches of data. A batch of data in this context pertains to the number of tensors being fed into the network for 1 epoch (forward pass and back-propagation). Different batch sizes will, therefore, impact the speed by which an epoch takes place. 

Figure \ref{fig:3dtensor} further depicts the tensor that is entered in the deep learning model. Each of the events is represented by a vector of size \textit{vocabularySize}. The number of preceding events considered is \textit{windowSize}, with the \textit{batchSize} as the number of records entering the input. The shape of the tensor can be expressed by $windowSize \times vocabularySize \times batchSize$. 

% TODO: \usepackage{graphicx} required
\begin{figure}
	\centering
	\includegraphics[width=.8\linewidth]{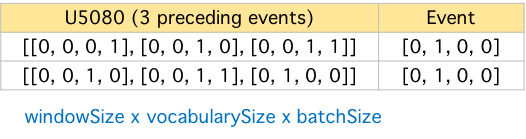}
	\caption{Input 3D tensor into the deep learning model.}
	\label{fig:3dtensor}
\end{figure}

The above designed tensor will be inputted to the neural network for training. As discussed in Section \ref{framework}, we use LSTM as the fundamental architecture for this study.
The deep learning model is depicted in Figure \ref{fig:lstmmodel} using Knime  data pipelining tool \cite{KNIMEOpe0:online} with two connected LSTM units and a dropout layer for regularization.

It should be noted that the objective is to detect the red team records (insider's threats) that are injected in the dataset. We select a particular user that has sufficient records for a deep learning network, without requiring vast amounts of time for training. Table \ref{tab:selectedUsers} provides an overview of some of the users in the dataset.
The list includes 7 users,  4 of them have red team records  and the remaining 3 capturing users that are considered normal. The explanation is as follows. 

\subsection{Training}
% TODO: \usepackage{graphicx} required
\begin{figure}[h]
	\centering
	\includegraphics[width=1\linewidth]{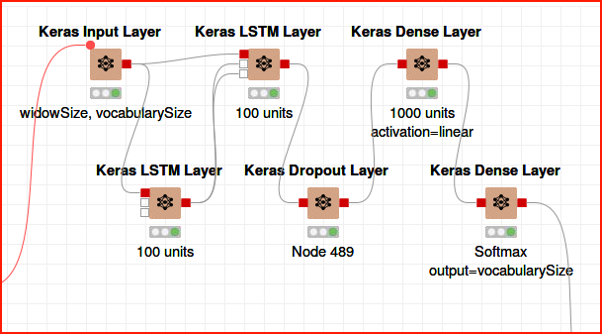}
	\caption{LSTM in sequence with dropout (model architecture).}
	\label{fig:lstmmodel}
\end{figure}
U66 has a very large quantity of authentications in the system, with a  large absolute number of red team records (118). However, the probability of getting a red team record by randomly picking one authentication event from U66 is extremely low at 0.0003\%. We choose not to experiment with this user given the very long training time that would be required. U1581 is at the other end of the continuum.  The percentage of red team records is very high (9.4\%) but the number of records is too low to warrant the use of deep learning.
\begin{table}[h!]
	\centering
	\begin{tabular}{||c|c|c|c||} 
		\hline
		User & Authentications & Red team & Probability  \\ [0.5ex] 
		\hline\hline
		U66 & 3,372,907 & 118 & 0.0003\% \\
		\hline
		U3005 & 44,150 & 36 & 0.1400\% \\
		\hline
		U737 & 55,549 & 33 & 0.0093\% \\
		\hline
		U1581 & 32 & 3 & 9.4\% \\
		\hline
		U748 & 63,370 & 0 & N/A \\
		\hline
		U4543 & 40,317 & 0 & N/A \\
		\hline
		U213 & 40,315 & 0 & N/A \\
		\hline
	\end{tabular}
	\vspace{3mm}
	\caption{Selected users with normal and red team records}
	\label{tab:selectedUsers}
\end{table}

U737 and U3005 appear more adequate for the experiments. The number of authentication records is suitable for deep learning, and the number of red team records suffices for partitioning the dataset in training and validation segments with the use of stratified sampling. 
\begin{table}[h!]
	\centering
	\begin{tabular}{||c|c|c|c||} 
		\hline
		User & Training (h:m:s) & Last batch accuracy & Red team \\ [0.5ex] 	
		\hline\hline
		U3005 & 0:04:11 & 50.00\% & Yes \\
		\hline
		U737 & 0:20:32 & 50.22\% & Yes \\
		\hline
		U748 &  0:34:45 & 49.34\% & Yes \\
		\hline
		U213 & 0:02:14 & 43.48\% & No \\
		\hline
		\hline
	\end{tabular}
	\vspace{3mm}
	\caption{Training results}
	\label{tab:trainingResults}
\end{table}

The normal users illustrated - U748, U4543, U213 - have sufficient records to have meaningful deep learning training, with no red team records present.

% TODO: \usepackage{graphicx} required
\begin{figure}
	\centering
	\includegraphics[width=1\linewidth]{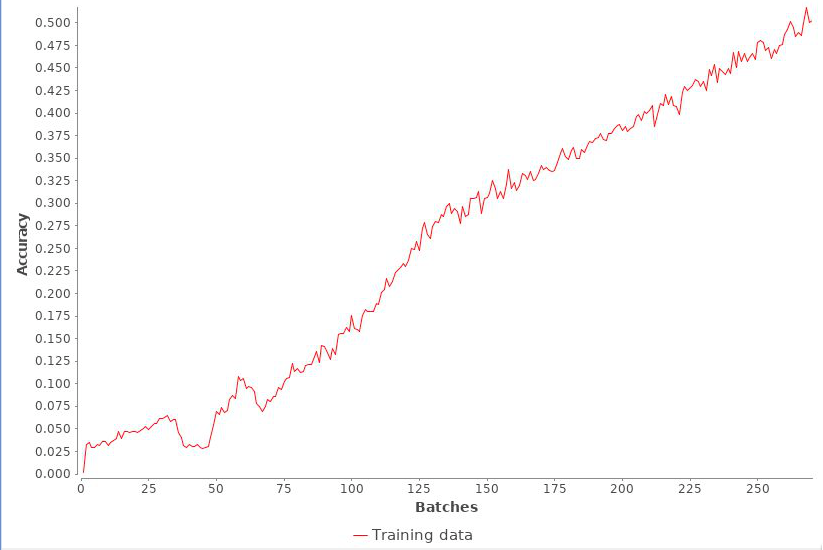}
	\caption{Training performance for user U737}
	\label{fig:u737training}
\end{figure}

Table \ref{tab:trainingResults} captures the key metrics for training the  selected users including the average training time for selected users and the last batch accuracy achieved.
The model was trained during 30 epochs with a batch size of 5,000 records. For user U737 -- with a total of 44,439 records for training -- this means  each epoch uses 9 batches for the training. 
The presence of red team records is displayed, although this information was not used in the training of the model.
Figure \ref{fig:u737training} illustrates that an accuracy of approximately 50.22\% was achieved after processing 270 batches during 20 minutes and 32 seconds. 

% TODO: \usepackage{graphicx} required
\begin{figure}
	\centering
	\includegraphics[width=1\linewidth]{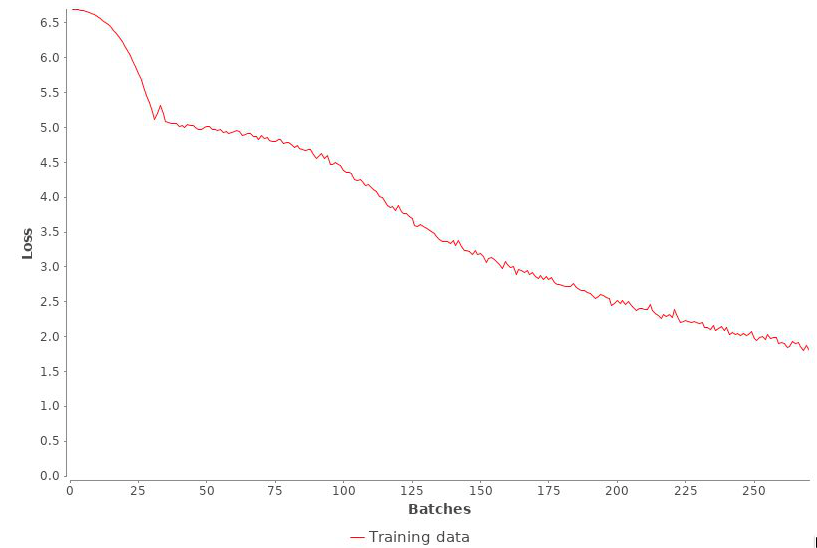}
	\caption{Loss function per batch over training time of U737}
	\label{fig:u737trainingloss}
\end{figure}
It is possible to observe how the learning took place by finding the minimum of the loss function shown in  equation \ref{lossFunction}. The calculated loss per batch is depicted in Figure \ref{fig:u737trainingloss}. 

It is also possible to compare and contrast the training time required for each user in the selection. Although U737 has approximately 30\% more records than U3005 (Table \ref{tab:selectedUsers}) the training time was five times longer (Table \ref{tab:trainingResults}). This can be explained by the vocabulary size. The sparse vector representing an event for U737 has a size of 812 (Table \ref{tab:tensorUsers}) compared with a size of 182 for U3005. This can also be observed in the training time for U213, which only took 2:14 minutes. 

The presence of red team records did not create any difference in the training phase of the analysis. This is due to the very few number of red team records that is present for some users as per Table \ref{tab:selectedUsers}.
\subsection{Prediction}
% TODO: \usepackage{graphicx} required
\begin{figure}
	\centering
	\includegraphics[width=1\linewidth]{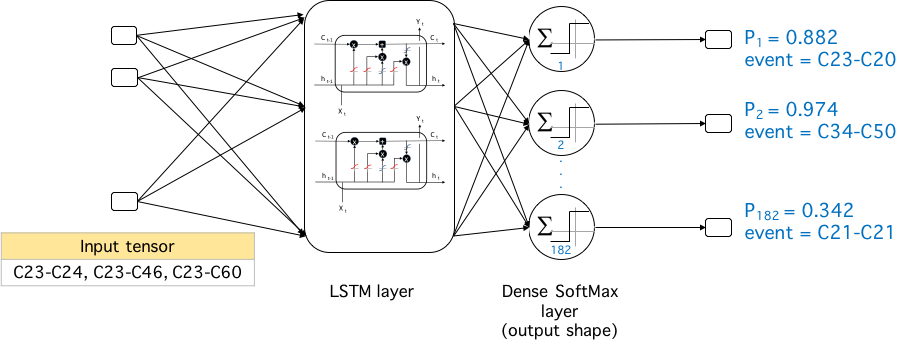}
	\caption{Last layer of the deep learning model for U3005 (dense softmax)}
	\label{fig:u737softmax}
\end{figure}
With the trained model, we proceed to predict the event (user behavior) for each of the sequences. We follow the typical percentages for training and testing data, i.e., 80\% for training and 20\% for prediction (testing) \cite{shmueli2017data}. As previously mentioned, the partitioning of data uses stratified sampling to ensure we have sufficient red team records in the prediction dataset (i.e., 20\% of the original dataset).

The output of the prediction yields the probability for each event in the vocabulary. In other words, for each potential event, a probability number is estimated by the model. We use this probability as a key metric in the decision making. For each input tensor composed of 30 events, we document the event with the highest probability as the predicted event. This dynamic is illustrated in Figure \ref{fig:u737softmax}. The input shape (i.e. the tensor created from the sequence) is fed into the LSTM layers. The final layer for the output of the model is a dense layer with as many neurons as the vocabulary size. In the case of the user U3005, the vocabulary size is 182 (as per Table \ref{tab:tensorUsers}). Thus, the deep learning model provides the different probabilities it assigned to each of the outputs. In the case of the illustration, the neuron 2 has the highest probability, and is therefore selected as the output, i.e., the second event in the vocabulary.

The above process is performed over the full test dataset (the 'new' data). In the case of U737, approximately 11,000 sequences are predicted. We can depict the results of the prediction using the ROC curve as explained in section \ref{approach}.

% TODO: \usepackage{graphicx} required
\begin{figure}
	\centering
	\includegraphics[width=1\linewidth]{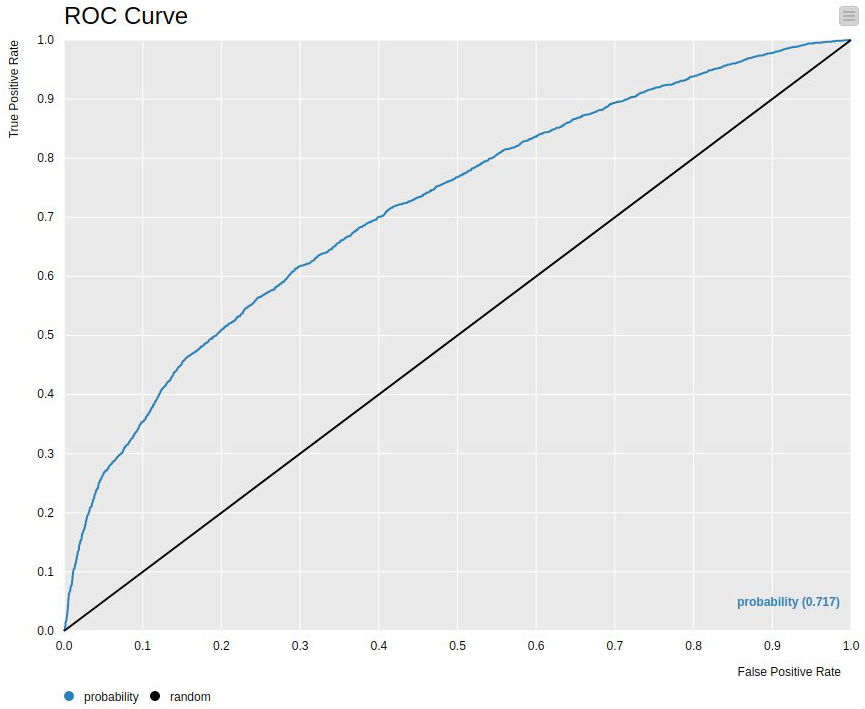}
	\caption{ROC curve for predicting one authentication event based on the preceding 30 authentications}
	\label{fig:u737roc}
\end{figure}

Figure \ref{fig:u737roc} shows the results. As can be observed, the deep learning model shows how the model is suitable for predicting the event when compared to a random draw. The Area Under the Curve (AUC) for U737 is 0.717. Similar results are obtained with other users as it is displayed in Table \ref{tab:predictionResults}.

The results obtained validate that an authentication event can be accurately predicted -- but how is this related with finding an insider's threat ('red team') taking place?

\begin{table}[h!]
	\centering
	\begin{tabular}{||c|c|c|c||} 
		\hline
		User & ROC's AUC & Threat in 10 lowest prob. records  \\ [0.5ex] 	
		\hline\hline
		U3005 & 0.761 & Yes \\
		\hline
		U737 & 0.717 & Yes \\
		\hline
		U748 & 0.687 & N/A \\
		\hline
		\hline
	\end{tabular}
	\vspace{3mm}
	\caption{Prediction results}
	\label{tab:predictionResults}
\end{table}

\subsection{Evaluation and adjustment}

Although the ability to predict the event has been validated, there is an additional process required for its application in the detection of the insiders' threat. 
We pay particular attention to the probability for predicted events. It is possible to segment the predictions as follows:

\subsubsection{High probability, correct prediction} This is a case to be expected if the model is performing well. A high probability for a predicted event should ideally result in the correct event predicted when compared with the actual event. From an insider's threat perspective, this scenario does not provide useful information.

\subsubsection{High probability, incorrect prediction} This is the case in which the model was quite certain that an event was going to happen, but the actual event was a different one. This may be a marker of an insider's threat taking place -- the model had high expectations for an event based on the historical data, but an anomaly took place. 

\subsubsection{Low probability, correct prediction} This is a situation that does not happen often. The deep learning model was not 'confident' in the prediction (as indicated by the low probability) but nonetheless it successfully predicted the event. From an insider's threat detection, this scenario is not relevant.

\subsubsection{Low probability, incorrect prediction} This situation can also be considered a marker for an insider's threat taking place. A low probability number conveys the message that the deep learning model was unable to place the event based on the learning from the training dataset. Since a red team record is an anomaly, it is to be expected that it'd have a low probability of detection and will end up being incorrect.

Figure \ref{fig:u3005lowprobabilityevents} lists the predictions under the third scenario articulated: low probability, incorrect predicted event. The records selected are the ones with the 10 lowest probabilities. The deep learning model failed to accurately predict any of those 10 events, but they contain 5 threat events. 

% TODO: \usepackage{graphicx} required
\begin{figure}
	\centering
	\includegraphics[width=1\linewidth]{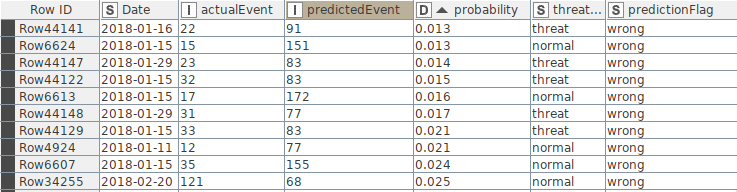}
	\caption{Predictions with low probability vs. ground truth for U3005}
	\label{fig:u3005lowprobabilityevents}
\end{figure}

The same analysis is performed with U737. The lowest probabilities are captured in Figure \ref{fig:u737lowprobabilityevents}. The first 10 lowest-probability records analyzed include 2 of the insider threat's records. This means that the heuristic used would enable a security analyst to limit further analysis to the lowest probability records for a user. For non-insider's threats, the security analyst will not find anything wrong, but the scrutiny on a very limited number of records may significantly improve the chances of catching the threat.

% TODO: \usepackage{graphicx} required
\begin{figure}
	\centering
	\includegraphics[width=1\linewidth]{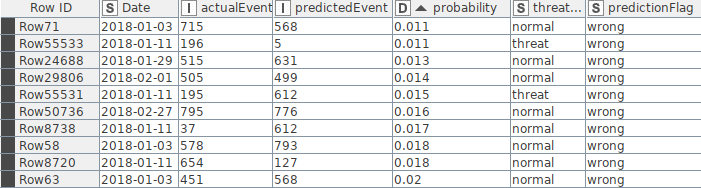}
	\caption{Predictions with low probability vs. ground truth for U737}
	\label{fig:u737lowprobabilityevents}
\end{figure}

%%%%%%%%%%%%%%%%%%%%%%%%%%%%%%%%% DISCUSSION %%%%%%%%%%%%%%%%%%%%%%%%%%%%%%%%%
% GUIDES:
% - Describe advantages and disadvantages of the proposed approach.
% - Explain shortcomings (e.g., not scalable, high complexity, heuristic-based, 
%   non accurate results).
% - Describe experiences you gained: i) the way you solved your practical problems; 
%   ii) specific advice for reducing the experimentation time; iii) attempts that were 
%   unsuccessful; iv) procedures for improving efficiency; etc.
%%%%%%%%%%%%%%%%%%%%%%%%%%%%%%%%%%%%%%%%%%%%%%%%%%%%%%%%%%%%%%%%%%%%%%%%%%%%%
\color{black}
\section{Conclusion}
\label{conclusion}
To conclude this work, it is relevant to review the principles and constraints identified in section \ref{approach}, as they holistically describe the strategy behind the experiments.

\paragraph{No normality baseline}
The experiments described in this study exist in alignment with the scenarios that would be encountered in practice. An existent insider's threat needs to be detected even without knowing what normality looks like for any given user. Thus, although the dataset does provide labeled data, the experiments do not use it for training purposes -- no normality signature exist. The process described is unsupervised in nature, using the ground truth just at the end of the exercise to assess results.
\paragraph{Finite training and detection time}
Throughout the experiments we have demonstrated that the training of the deep learning model -- the most resource-intensive process in the architecture -- is performed in an efficient manner. For users with low vocabulary sizes, the training takes place in under 5 minutes.
\paragraph{Very large datasets}
The dataset used for the experiments is very large, around 1.6 billion records in total. To best utilize the available resources, each user is analyzed independently -- achieving a good balance between sufficient data for deep learning and fast training and detection.
\paragraph{Probability-based}
The approach designed uses the probabilities estimated by the model for each of the predicted events. Using this information, it is possible to decrease the search space for the analysis -- a security analyst would need to analyze a small set of records in order to find an insider's threat taking place. The experiments demonstrated how this work for two specific cases of U3005 and U737. The heuristic employed was effective for the task: analysis of the high probability events that were wrongly predicted, as well as low probability events estimated by the model. 
\paragraph{Inherent characteristics in the data}
The authentications information found in the dataset follow a sequence in time. Using the historical information (the preceding authentication events) it was possible to predict the next event, and use the associated probability information as markers of potential insider's threats. Using LSTM proved to be a suitable architecture for detection, since it takes advantage of the sequenced nature of the data to find the next event expected.\\

There are multiple new research streams that may prove useful for researchers and applicable to practitioners. We describe two of them:
\paragraph{Use of word embeddings} the encoding process assigned an integer number (an 'index') to the events under analysis. This was a required step in order to feed the data to a computational model. However, each index exist independently of the other ones. I.e., event "4" and event "16" may be related in some way, but the encoding did not reflect that relationship. A potential approach that could assign meaning to the encoding is the use of word embeddings. Algorithms such as Word2Vec can assign meaningful numeric vectors to each event, which would open a very rich discussion on inter-event relationships and how they may add value to the insider's threat detection processes.
\paragraph{Transfer learning} The approach taken in this research used Keras models trained by user. It is possible to contemplate that a Keras model trained on one user may be relevant for another user. This is only feasible if the encoding is common to all users: i.e. event "4" for one user is the same event "4" for another user. Under this scenario, it is possible to think that insights for a sequence of events may be usable from user to user. The immediate impact of this approach is that the vocabulary size would become significantly larger in order to include all events for any user -- impacting significantly some of the principles discussed previously. \\

The detection of an insider's threat taking place is a rich ground for the use of deep learning. The experiments described here provide a glimpse of the value that machine learning brings to the detection of the insider's threat.
\color{black}
%\input{7.Conclusion}
%\input{8.IEEEconference_Sample}

%------------------------------------------ END SECTIONS ------------------------------------------

%\begin{IEEEkeywords}
%component, formatting, style, styling, insert
%\end{IEEEkeywords}

\bibliographystyle{plain}
\bibliography{library}

\end{document}